# Focusing Metasurfaces of (Un)equal Power Allocations for Wireless Power Transfer

Andi Ding, *Graduate Student Member*, *IEEE*, Yee Hui Lee, *Senior Member*, *IEEE*, Eng Leong Tan, *Senior Member*, *IEEE,* Yufei Zhao, Yanqiu Jia, *Graduate Student Member, IEEE*, Yong Liang Guan, *Senior Member*, *IEEE*, Theng Huat Gan, *Senior Member*, *IEEE,* Cedric W. L. Lee, *Senior Member*, *IEEE*

*Abstract*— Focusing metasurfaces (MTSs) tailored for different power allocations in wireless power transfer (WPT) system are proposed in this letter. The designed metasurface unit cells ensure that the phase shift can cover over a $2\pi$ span with high transmittance. Based on near-field focusing theory, an adapted formula is employed to guide the phase distribution for compensating incident waves. Three MTSs, each with dimensions of 190×190 mm and comprising 19×19 unit cells, are constructed to achieve dual-polarized two foci with 1:1, 2:1, and 3:1 power allocations, yielding maximum focusing efficiencies of 71.6%, 65.2%, and 57.5%, respectively. The first two MTSs are fabricated and tested, demonstrating minimal −3 dB depth of focus (DOF). Results are aligned with theoretical predictions. These designs aim to facilitate power transfer to different systems based on their specific requirements in an internet of things (IoT) environment.

## I. Introduction

Wireless power transfer (WPT) for electromagnetic wave in the near field has gained increasing attention with the proliferation of internet of things (IoT) applications. Various WPT approaches, including radio-frequency identification (RFID) [1] and beamforming [2]-[3], have been extensively explored. Among these, focusing metasurfaces (MTSs), evolved from traditional dielectric lenses [4][5], have garnered significant interest due to their compact flat structures and multifunctional capabilities [6].

Recent research aim to enhance MTS designs for effective focusing performance, with transmissive MTS typically incorporating unit cells capable of large phase shifts and high transmission responses. Consequently, multi-layer structures are widely adopted to increase phase variation across elements. For example, [7] proposed a 3-bit coding MTS consisting of three-layer structure to improve resolution efficiency, while [8] utilized four identical layers to enhance dual-polarization transmission efficiency. However, many power transfer applications, particularly in IoT scenarios, require a feed capable of powering multiple devices simultaneously, a need unmet by these traditional designs [9].

Multi-foci MTS designs address this challenge [10][11], in which the phase compensation principle needs to be refined to facilitate multiple foci. [12] proposed multi-foci formulation to generate two equal foci. However, the resulting efficiency remains suboptimal for practical applications. More advanced implementations, such as [13], achieve up to 23 equal foci on a spherical surface using an adapted Friis formula, but overall efficiency remains below 50%. Works such as [14] and [15] have enhanced efficiency through optimization of spatial correlation between unit cell locations, albeit with a primary focus on achieving equal-power foci. To the best of our knowledge, little research has explored unequal power allocation among multiple foci, a critical requirement for real-world applications where different systems have distinct power demands.

To address this gap, this letter proposes three focusing MTSs designs capable of dual-polarized dual-foci 1:1, 2:1 and 3:1 power allocations with a maximum focusing efficiency of 71.6%, 65.2% and 57.5%, respectively. These MTSs, each measuring 190×190 mm and consisting of 19×19 unit cells, employ four-layer identical patterns that achieve a phase shift exceeding $2\pi$ while maintaining high transmittance.

## II. Focusing MTSs Design and Analysis

### A. Unit Cell Design

The configuration of the proposed unit cell for the MTS is depicted in Fig. 1(a) and (b). Each square unit has a periodicity of 10 mm and is closely packed to form an array. Outer loop with small width reduces interferences from neighboring unit cells. The radius of the central circle, tunable between 0.8 mm and 4.5 mm, determines transmission phase response at the target frequency of 10 GHz. A square hole is introduced at the center to enhance transmittance while maintaining phase stability across different oblique incident angles. The unit cells consist of four metallic layers printed on three F4BM dielectric layers ($\varepsilon_r =$ 2.2 and tan $\delta$ = 0.0015), each with a thickness of 3 mm.

The proposed unit cell is simulated using CST Suite Studio with Floquet port setup. The transmission magnitudes of unit cells with $r$ = 0.8, 2.5, 3.5, 4, 4.5 mm are shown in Fig. 1(c), demonstrating that resonant frequencies vary with radius of the unit cell, yielding a shared operational bandwidth of 9.3-10.6 GHz (13%). This bandwidth provides tolerance for minor frequency shifts during fabrication and application. Fig. 1(d) illustrates transmission magnitude and phase of unit cells with different radii for dual-polarized operation at 0° and 30° incidence at 10 GHz. It is observed that all curves related to magnitude are above −1 dB and those related to phase can cover $2\pi$ phase shift, confirming minimal transmission loss and high angular stability.

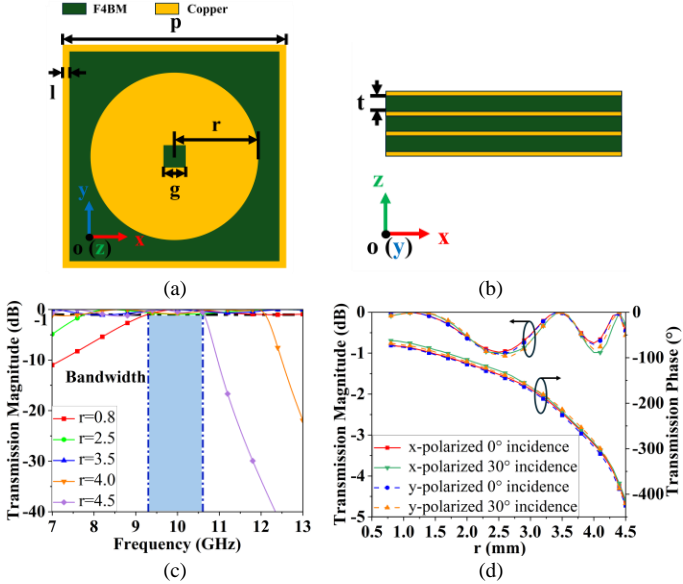

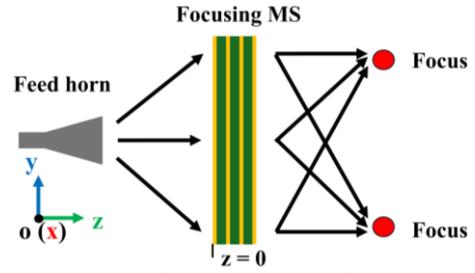

Fig. 2. Working principle of a focusing MTS with 2 foci.

Fig. 1. (a) Top view and (b) side view of the proposed unit cell. $p$ = 10 mm, $l$ = 0.2 mm, $g$ = 0.5 mm, $t$ = 3 mm. (c) Transmission magnitude of unit cells with $r$ = 0.8, 2.5, 3.5, 4, 4.5 mm, respectively. (d) Transmission magnitude and phase of unit cells with different radii for dual polarization at 0º and 30º incidence at 10 GHz.

B. Phase Distribution Principle

In order to transform the spherical wave emitted by the horn feed into a focused beam in the near field, the MTS is designed to compensate for phase differences using unit cells with tailored phase responses. A quantized phase gradient approach simplifies calculations by considering discrete phase compensations rather than a continuous function [16]. As shown in the conceptual diagram of the focusing MTS in Fig. 2, the total phase compensation for a given unit cell can be divided into two parts, which can be expressed as

$$\varphi_{i,j} = \varphi_{i,j|T} + \varphi_{i,j|F} \quad (1)$$

where $\varphi_{i,j}$ represents the total phase compensation for the unit cell of the $i^{th}$ row and the $j^{th}$ column, $\varphi_{i,j|T}$ and $\varphi_{i,j|F}$ represent the phase compensation from the feed to the unit cell and from the unit cell to the foci, respectively. Then for the first part of (1), it is well known that

$$\varphi_{i,j|T} = k \cdot d_{i,j|T} \quad (2)$$

where $k$ is the free-space wavenumber, and $d_{i,j|T}$ denotes the distance from the phase center of the feed to the center of the target unit.

Inspired by Gerchberg-Saxton Weighting (GSW) algorithm [17], an adapted multi-foci formula is proposed incorporating weighting factors before each component to acquire the second part of the phase compensation. The formula of forming $n$ independent foci can be written as

$$\varphi_{i,j|F} = \arg\left[\sum_1^n w_n \cdot \exp(jk \cdot d_{i,j|Fn})\right] \quad (3)$$

where $w_n$ is the weight for the $n^{th}$ focus, and $d_{i,j|Fn}$ denotes the distance from the $n^{th}$ focus to the target unit. Various power allocations can be obtained by tuning the weights assigned to the foci based on this formula. In this letter, we will focus on the two-foci case with different locations and power ratios.

C. Different power ratios

Integrated with the proposed unit cells and focusing formula, MTSs of power allocations of 1:1, 2:1 and 3:1 for the two-foci case, will be analyzed in detail. Each MTS comprises 19 unit cells along both x- and y-axes, forming an aperture of 190× 190 mm$^2$, which ensures sufficient transmitting power. The focal length to diameter ratio ($F/D$) is then maintained at 0.55 to achieve high transmitting efficiency. For clarification, the first layer of the MTS is on the $z$ = 0 plane in Fig. 2. Both weighting factors are then set to 1 for the equal foci case in (3) and the target foci are positioned at ($\pm60$, 0, 90) mm for x-polarization. Phase distribution for the 1:1 power allocation case is then be calculated and plotted in Fig. 3(a). The simulated normalized E-field intensities for x-polarization on the *xoz* and *xoy* planes at 10 GHz are plotted in Fig. 4(a). It is observed that the foci and their −3 dB region are symmetric about the *x*-axis. Additionally, the −3 dB depth of focus (DOF: length of the *z*-axis projection) and width of focus (WOF: maximum radius on the *xoy* plane) are 45 mm and 24 mm, respectively. To better quantify the focusing effect, we define the focusing efficiency $\eta_F$ as the ratio of focused power $P_F$ to the incident power on the MTS, $P_{IM}$ [18]. Poynting vector is used to determine the field power intensity, so $\eta_F$ can be expressed as

$$\eta_F = \frac{P_F}{P_{IM}} = \frac{\iint_{S_F} Re(\vec{E} \times \vec{H}) \cdot d\vec{S}}{\iint_{S_M} Re(\vec{E} \times \vec{H}) \cdot d\vec{S}} \quad (4)$$

where $S_F$ is the total −3 dB region of the focal points, and $S_{IM}$ denotes the aperture of the MTS. Then the focusing efficiency for x-polarized equal-foci case is calculated as 69.9%. Similarly, the normalized E-field intensities for y-polarization on the *yoz* plane and the *xoy* plane at 10 GHz is obtained by rotating the horn antenna and is plotted in Fig. 4(b), showing little change in DOF and WOF. The corresponding focusing efficiency is 71.6%.

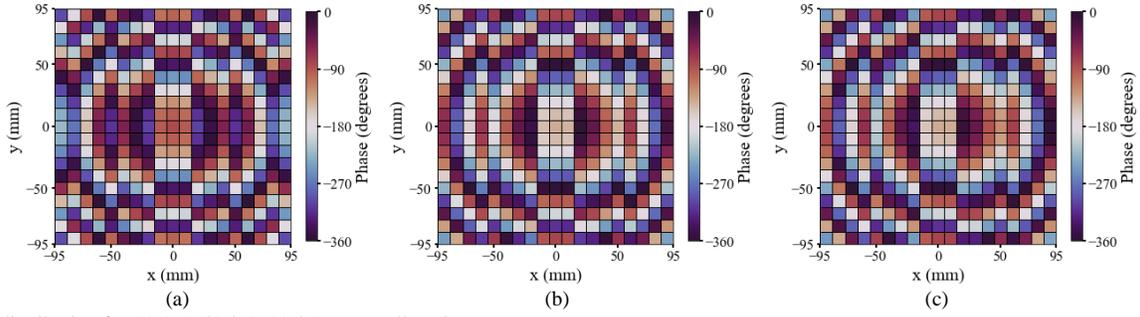

Fig. 3. Phase distribution for (a) 1:1, (b) 2:1, (c) 3:1 power allocation cases.

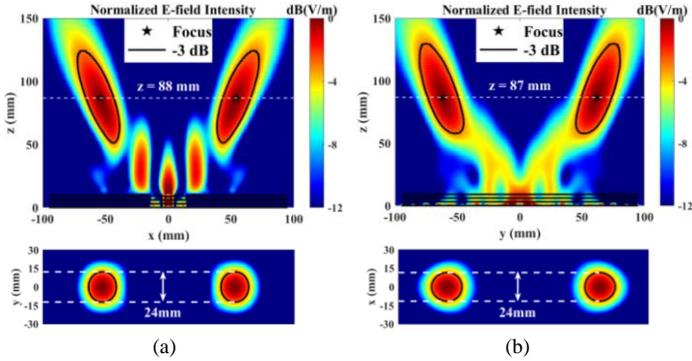

Fig. 4. 1:1 power allocation case: Normalized E-field intensities on the *xoz/yoz* and *xoy* plane for (a) x-polarization and (b) y-polarization at 10 GHz. (Max. magnitudes are 50.9 and 50.6 dBV/m for (a) and (b), respectively.)

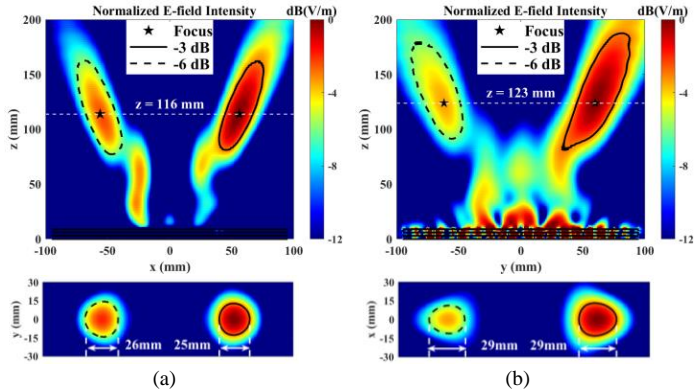

Fig. 5. 2:1 power allocation case: Normalized E-field intensities on the *xoz/yoz* and *xoy* plane for (a) x-polarization and (b) y-polarization at 10 GHz. (Max. magnitudes of two foci are 51.5 and 48.9 dBV/m for (a), 51.3 and 48.4 dBV/m for (b), respectively.)

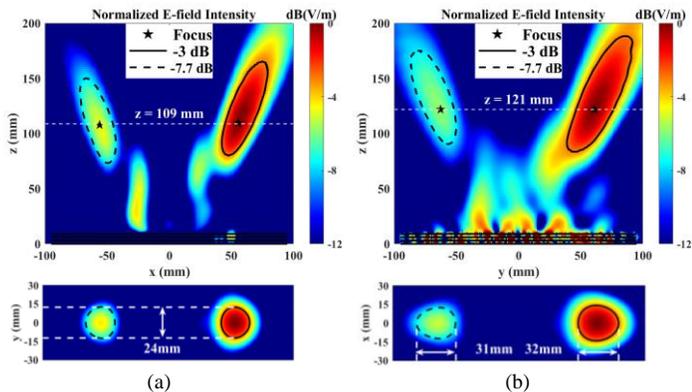

Fig. 6. 3:1 power allocation case: Normalized E-field intensities on the *xoz/yoz* and *xoy* plane for (a) x-polarization and (b) y-polarization at 10 GHz. (Max. magnitudes of two foci are 51.1 and 46.8 dBV/m for (a), 51.0 and 46.2 dBV/m for (b), respectively.)

The amplitude ratio of foci is sensitive to the change of discrete phases across the MTS. Consequently, the weighting factors in (3) are slightly changed to 1.04 and 0.96 for the 2:1 power allocation case. Target foci are pushed further towards the positive direction of the z-axis at ($\pm$60, 0, 120) mm for x-polarization and phase distribution is plotted in Fig. 3(b). The normalized E-field intensities on the *xoz/yoz* and *xoy* planes for x- and y-polarizations at 10 GHz are plotted in Fig. 5(a) and (b), respectively. Although a 3 dB magnitude difference is observed between the two foci, their own −3 dB region are still symmetric. The −3 dB DOF and WOF are 51 mm and 26 mm for x-polarization with a focusing efficiency of 65.2% and are 54 mm and 29 mm for y-polarization with a focusing efficiency of 62.3%, respectively. Similarly for 3:1 power allocation case, weighting factors are changed to 1.1 and 0.9, and the corresponding phase distribution is plotted in Fig. 3(c). The normalized E-field intensities on the *xoz/yoz* and *xoy* planes for x- and y-polarizations at 10 GHz are plotted in Fig. 6(a) and (b), respectively. A 4.7 dB difference is spotted between two peaks, and the −3 dB regions remain symmetric. The −3 dB DOF and WOF are 48 mm and 24 mm for x-polarization with a focusing efficiency of 57.3% and are 60 mm and 32 mm for y-polarization with a focusing efficiency of 52.1%, respectively. Although the differences in magnitude become larger for the above three scenarios, servicing regions of the foci remain unchanged, suggesting the stability of this series of designs. When the power ratio is 3:1, the focusing efficiencies are decreased for both polarizations due to the interference caused by unequal power ratio for the two foci. Still, it can be concluded that high-efficiency two-foci power allocations are realized, and target locations of the foci can be directly controlled by the formula (3).

## III. FABRICATION AND MEASUREMENTS

The MTSs for 1:1 and 2:1 power ratios are fabricated to validate the proposed designs. The top surface of the prototypes is shown in Fig. 7(a). Eight via holes are drilled into each substrate board, and nylon screws are used to fasten them together. The MTS and feed horn are then fixed on a foam

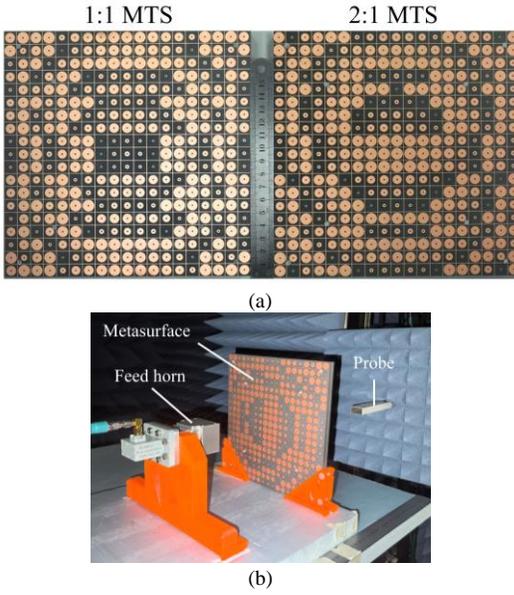

Fig. 7. Photos of (a) fabricated prototype of 2 MTSs and (b) measurement setup in anechoic chamber.

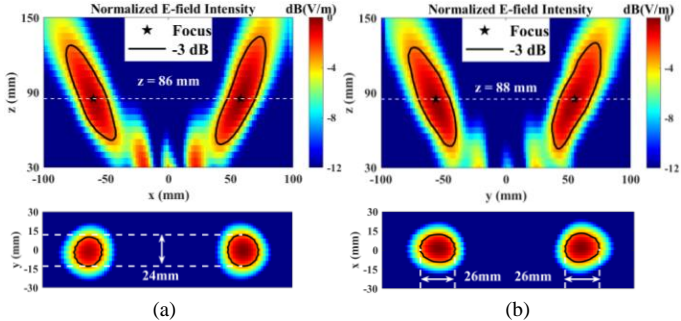

Fig. 8. Measured 1:1 power allocation case: Normalized E-field intensities on the *xoz/yoz* and *xoy* plane for (a) x-polarization and (b) y-polarization at 10 GHz.

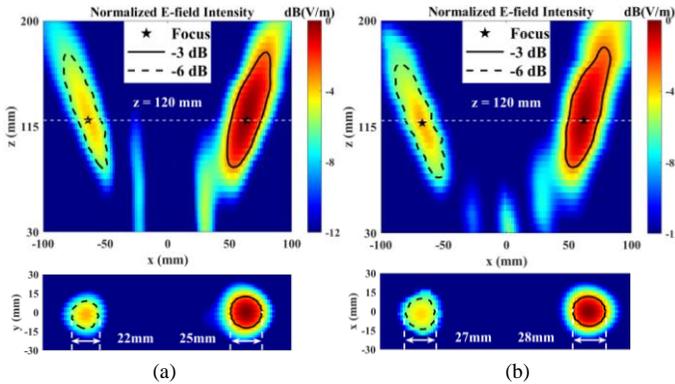

Fig. 9. Measured 2:1 power allocation case: Normalized E-field intensities on the *xoz/yoz* and *xoy* plane for (a) x-polarization and (b) y-polarization at 10 GHz.

TABLE I
COMPARISON OF THE PROPOSED METASURFACES WITH OTHER DESIGNS IN LITERATURE

| Ref. | Freq. (GHz) | Diameter ($\lambda$*) | Max. focusing efficiency | Polarization | PA** |
|---|---|---|---|---|---|
| [11] | 13 | 5.1 | 40% | Single-Pol. | 1:1 |
| [13] | 10 | 13 | 68.3% | Single-Pol. | 1:1 |
| [14] | 5.8 | 8.8 | 70% | Single-Pol. | 1:1 |
| **Ours** | 10 | 6.3 | 71.6% | **Dual-Pol.** | **1:1** |
| | | | 65.2% | **Dual-Pol.** | **2:1** |
| | | | 57.3% | **Dual-Pol.** | **3:1** |

* $\lambda$ is the wavelength corresponding to the center frequency
** PA is power allocation for two foci

board using plastic supports in the anechoic chamber with a distance of 105 mm between them, as shown in Fig. 7(b). The probe is placed behind the MTS and is remotely controlled by a computer to perform the measurement.

The dual-polarized normalized E-field intensities on the *xoz/yoz* and *xoy* planes of the first MTS are measured, and shown in Fig. 8(a)-(b). For the 1:1 power allocation case, the scanning plane dimensions for the *xoz/yoz* and *xoy* planes are 200×120 mm² and 200×60 mm², respectively. The starting position on the z-axis is set to 30 mm to avoid collision during testing. The measured −3 dB DOF and WOF are 47 mm and 24 mm for x-polarization and are 49 mm and 26 mm for y-polarization, respectively. Similarly for the 2:1 power allocation case, the measured dual-polarized normalized E-field intensities on the *xoz/yoz* and *xoy* planes are shown in Fig. 9(a)-(b). The *xoz/yoz* scanning plane dimensions are increased to 200 ×170 mm² to cover the focus region. The measured −3 dB DOF and WOF are 54 mm and 25 mm for x-polarization and are 54 mm and 28 mm for y-polarization, respectively. Both groups of results align with the simulated ones, and the differences observed are estimated to be caused by the interference in power division and occlusion from the supports used to perform the measurement.

Finally, a comparison between our proposed MTSs and other published designs in the literature is done and shown in Table I. Although there are many excellent works in this field, we have only selected articles that fully align with our research topic. It is clear that our design with a 1:1 power allocation achieves the highest focusing efficiency with a relatively small aperture size. More importantly, our designs also offer dual-polarized MTSs of unequal (2:1 and 3:1) power ratios to cater to needs in various application requirements.

## IV. CONCLUSION

Focusing MTSs are proposed in this letter to realize dual-polarized two foci with 1:1, 2:1 and 3:1 power allocations at 10 GHz in near-field WPT applications. Each MTS has 19 unit cells in each of the 2 dimensions, forming an aperture of 190×190 mm that ensures high transmitting power. They also feature unit cells engineered for high transmitting efficiency and broad $2\pi$ phase coverage, ensuring robust dual-polarized performances. Multi-foci formula is adapted with weighting factors to enable efficient asymmetric power allocation without increasing interference. The maximum focusing efficiencies for the three power ratio scenarios are 71.6%, 65.2% and 57.3%, respectively, with small DOFs and WOFs. Moreover, two of these designs with 1:1 and 2:1 power allocations are fabricated and tested in an anechoic chamber. Experimental validations confirm the practicality of these designs, highlighting their potential for powering multiple distinct IoT devices in complex environments.